\newif\ifpnas

\pnasfalse

\ifpnas

\documentclass[9pt,twocolumn,twoside]{pnas-new}

\templatetype{pnasresearcharticle} 

\usepackage{dsfont}
\definecolor{myBlue}{rgb}{0.25,0.8,0.98}

\def\s{\sigma}
\DeclareMathOperator{\tr}{Tr}
\renewcommand{\epsilon}{\varepsilon}
\def\@fnsymbol#1{\ensuremath{\ifcase#1\or *\or \dagger\or \ddagger\or
   \mathsection\or \mathparagraph\or \|\or **\or \dagger\dagger
   \or \ddagger\ddagger \else\@ctrerr\fi}}

\title{Classification of topological phonons in linear mechanical metamaterials}

\author[a,1]{Roman S\"usstrunk}
\author[a]{Sebastian D.~Huber}
\affil[a]{Institute for Theoretical Physics, ETH Zurich, 8093 Z\"urich, Switzerland}

\leadauthor{S\"usstrunk} 

\significancestatement{A mechanical metamaterial is an engineered material that is characterized by properties that go beyond the properties of its microscopic building blocks. Specifically, in topological metamaterials one makes use of surface our boundary modes that are stable against imperfections or environmental influences and hence constitute reliable building blocks for various applications. In our work we provide a classification scheme of possible topological metamaterials and an extensive number of examples illustrating this scheme. Our classification can serve as an important blueprint for many future applications that target such stable boundary modes for engineering purposes.}

\authorcontributions{Both authors contributed equally to this work.}
\authordeclaration{The authors declare no conflict of interest.}
\correspondingauthor{\textsuperscript{1}To whom correspondence should be addressed. E-mail: suesstrunk@itp.phys.ethz.ch}

\keywords{topological matter $|$ mechanical metamaterials $|$ adaptive materials}

\begin{abstract}
Topological phononic crystals, alike their electronic counterparts, are characterized by a bulk-edge correspondence where the interior of a material dictates the existence of stable surface or boundary modes. In the mechanical setup, such surface modes can be used for various applications such as wave-guiding, vibration isolation, or the design of static properties such as stable floppy modes where parts of a system move freely. Here, we provide a classification scheme of topological phonons based on local symmetries. We import and adapt the classification of non-interacting electron systems and embed it into the mechanical setup. Moreover, we provide an extensive set of examples that illustrate our scheme and can be used to generate new models in unexplored symmetry classes. Our works unifies the vast recent literature on topological phonons and paves the way to future applications of topological surface modes in mechanical metamaterials.
\end{abstract}

\dates{This manuscript was compiled on \today}
\doi{\url{www.pnas.org/cgi/doi/10.1073/pnas.XXXXXXXXXX}}

\begin{document}

\verticaladjustment{-2pt}

\maketitle
\thispagestyle{firststyle}
\ifthenelse{\boolean{shortarticle}}{\ifthenelse{\boolean{singlecolumn}}{\abscontentformatted}{\abscontent}}{}


\else

\documentclass[
reprint,
showpacs,
nofootinbib,
amsmath,amssymb,
aps,
pra,
floatfix,
]{revtex4-1}

\usepackage{dsfont}
\usepackage{graphicx}
\usepackage{xcolor}
\definecolor{myBlue}{rgb}{0.25,0.8,0.98}
\usepackage{amsmath}

\usepackage{hyperref}
\hypersetup{
  colorlinks = true, linkcolor = brown, citecolor = purple
}

\def\s{\sigma}
\DeclareMathOperator{\tr}{Tr}
\renewcommand{\epsilon}{\varepsilon}

\begin{document}

\title{Classification of topological phonons in linear mechanical metamaterials}
\begin{abstract}
Topological phononic crystals, alike their electronic counterparts, are characterized by a bulk-edge correspondence where the interior of a material dictates the existence of stable surface or boundary modes. In the mechanical setup, such surface modes can be used for various applications such as wave-guiding, vibration isolation, or the design of static properties such as stable floppy modes where parts of a system move freely. Here, we provide a classification scheme of topological phonons based on local symmetries. We import and adapt the classification of non-interacting electron systems and embed it into the mechanical setup. Moreover, we provide an extensive set of examples that illustrate our scheme and can be used to generate new models in unexplored symmetry classes. Our works unifies the vast recent literature on topological phonons and paves the way to future applications of topological surface modes in mechanical metamaterials.
\end{abstract}

\author{Roman S\"usstrunk}
\author{Sebastian D. Huber}
\affiliation{Institute for Theoretical Physics, ETH Zurich, 8093 Z\"urich, Switzerland}

\maketitle

\fi

\ifpnas \dropcap{M}\else M%
\fi%
echanical metamaterials derive their properties not from their microscopic composition but rather through a clever engineering of their structure at larger scales \cite{Cummer16}. Various design principles have been put forward and successfully applied in the past. Examples range from periodic modifications leading to band-gaps via Bragg scattering \cite{Kushwaha93} to the use of local resonances \cite{Liu00} to achieve sub-wavelength functionalities. Recently, the concept of ``band-topology'' emerged as a new design principle for mechanical metamaterials \cite{Prodan09, Kane13, Chen14, Chen15, Paulose15, Paulose15a, Xiao15, Susstrunk15, Nash15, He15, Meeussen16}. Colloquially speaking, a system with a topological phonon band-structure will posses mechanical modes bound to surfaces or lattice defects that are immune to a large class of perturbations. If the targeted purpose of a metamaterial is encoded in such a topologically protected mode, its functioning will be largely independent of production imperfections or environmental influences. 

The introduction of topology to the field of mechanical metamaterials was largely motivated by its successful application to the description of electrons in solids \cite{Hasan10}. One of the key elements in the understanding of these electronic systems was the classification of different topological phases according to their symmetry properties \cite{Kitaev09,  Ryu10}\nocite{Schnyder08}. While over the last years numerous proposals \cite{Prodan09, Berg11, Kane13, Po14, Yang15, Kariyado15, Kariyado15a, Yang16, Vitelli14, Wang15a, Peano15, Rocklin15, Rocklin15a, Sussman15, Lubensky15, Pal16, Salerno16, Khanikaev15, Mousavi15, Xiao15b, Ni15, Wang15b} and several experiments \cite{Chen14, Chen15, Paulose15, Paulose15a, Xiao15, Susstrunk15, Nash15, He15, Meeussen16} were put forward promoting mechanical topological metamaterials, a complete classification of linear topological phonons is missing to date. In this report we intend to fill in this gap. 

At first sight, the dynamics in classical mechanics seems to be rather different from quantum-mechanical electron systems. Our approach is therefore to map the first to the second problem \cite{Kariyado15}. This, in principle, allows us to import the classification \cite{Kitaev09, Ryu10} from the description of electronic systems. However, a bare import of this classification is not doing justice to the rich structure mechanical systems posses by themselves.

We can categorize mechanical metamaterials by two independent properties. First, the targeted functionality can either be at zero or at finite frequencies. Zero-frequency modes  define structural properties such as mechanisms where parts of a material move freely \cite{Kane13, Paulose15}. The dual partners of freely moving parts are states of self stress \cite{Lubensky15}, where external loads on a material can be absorbed in the region of a topological boundary mode. Defining such details of the load bearing properties of a material are relevant both for smart adaptive materials \cite{Paulose15a} as well as for civil engineering applications. The design of finite-frequency properties, on the other hand, constitutes a quite different field of research. Here, the goals are to control the propagation, reflection or absorption of mechanical vibrations. This includes, e.g., wave-guiding, acoustic cloaking, or vibration isolation ranging from the seismic all the way to the radio-frequency scale.

A second important separation into two distinct classes of materials arises from the presence or absence of non-reciprocal elements \cite{Fleury14}. Generically, non-dissipative mechanical properties are invariant under the reversal of the arrow of time. Non-reciprocal elements, however, transmit waves asymmetrically between different points in space. The absence of time-reversal symmetry allows for a topological invariant, the Chern number, which encodes chiral, or uni-directional wave-propagation. We will see that these two attributes: static vs. dynamic and reciprocal vs. non-reciprocal will be key to understand how the electronic classification is naturally modified for mechanical systems.  

Before we embark on the development of the framework needed for our classification let us state our goals more precisely. Our aim is to import and adapt the classification of non-interacting electron systems according to their local symmetries $\mathcal T$, $\mathcal C$, and $\mathcal S=\mathcal T \circ \mathcal C$, i.e., time-reversal, charge-conjugation, and their combination, respectively \cite{Kitaev09, Ryu10}. Clearly, we will have to specify the role of these symmetries in mechanical systems. Moreover, we only cover the ``strong'' indices which do not rely on any spatial symmetries. The extension to weak indices, arising from a stacking of lower-dimensional systems carrying strong indices, is straight-forward \cite{Kane13}. Finally, there are many recent developments dealing with topological phases stabilized by spatial properties \cite{Teo08, Fu11, Hsieh12, Xu12, Liu14} such as point group symmetries. While such spatial symmetries are more easily broken by disorder, the required ingredients might be very well tailored to the mechanical setup \cite{Alexandradinata14, Alexandradinata15}. 

The remainder of this paper is organized as follows: We start by developing a framework to map classical problems to an equation that formally looks like a Schr\"odinger equation of a quantum mechanical problem. We then introduce the three symmetries $\mathcal T$, $\mathcal C$, and $\mathcal S$ and discuss their appearance in mechanical problems before we provide the sought classification. Finally, an extensive example section serves two purposes: We illustrate and apply our approach. Moreover, we show a way how to construct new symmetry classes from generic building blocks. 

\section*{Models and theoretical framework}

In this manuscript we aim at characterizing discrete systems of undamped, linear mechanical oscillators. While this setup is directly relevant for simple mass-spring systems \cite{Susstrunk15} or magnetically coupled gyroscopes \cite{Nash15}, the scope here is actually considerably broader. Any system that can be reliably reduced to a discrete linear model is amenable to our treatment. This includes one \cite{Xiao15b}, two \cite{Khanikaev15,Ni15,Mousavi15,Yang15}, or three-dimensional \cite{Xiao15b} systems made from continuous media, where a targeted micro-structuring enables the description in terms of a discrete model. Once we deal with a discrete model, we have a direct way to import the methods known from electronic topological insulators. In order to establish this bridge we now introduce a formal mapping of a classical system of coupled oscillators to a tight-binding hopping problem of electrons in solids.

We start with the equations of motion of a generalized mass-spring model given by
\begin{equation}\label{eq:eom}
	\ddot x_i(t) = \sum_{j=1}^N\left[-D_{ij} x_j(t) + \Gamma_{ij} \dot x_j(t)\right]\,.
\end{equation}
Here, $t$ denotes time, $x_i(t)\in\mathds{R}$ one of the $N$ independent displacements, and $\dot {x}_i (t)$ its time derivative. The mass terms are absorbed into the real and constant coupling elements $D_{ij}$ and $\Gamma_{ij}$. The entries $D_{ij}$ can be thought of as springs coupling different degrees of freedom, and $\Gamma_{ij}$ arise from velocity dependent forces. Note, that a non-zero $\Gamma$ implies terms formally equivalent to the Lorentz force of charged particles in a magnetic field and hence arise only in metamaterials with non-reciprocal elements. In addition to constant coupling elements in \ref{eq:eom}, one can also consider periodically driven systems. Such driven system can be cast into our framework by a suitable (Magnus) expansion of the corresponding Floquet operator \cite{Floquet83, Lindner11b, Salerno16}.

We aim at rewriting equation~\ref{eq:eom}, in the form of a Schr\"odinger equation, or rather as a hermitian eigenvalue problem. Therefore, we need the system to be conservative (non-dissipative). This is achieved by requiring $D$ to be symmetric positive-definite and $\Gamma$ to be skew-symmetric.\footnote{
 Such matrices can, e.g., be obtained from a system with Lagrangian
$
	\mathcal{L} = \sum_{i,j} \dot{x}_i A_{ij} \dot{x}_j+x_iB_{ij} \dot{x}_j-x_i C_{ij}x_j\,,\qquad A_{ij},\,B_{ij},\,C_{ij}\in\mathds{R}\,,
$
given that $D=(A+A^T)^{-1} (C+C^T)>0$.}

An eigenvalue problem emerges from equation~\ref{eq:eom} via the ansatz $x_i(t)=e^{-i \lambda t} x_i(0)$:
\begin{equation}\label{eq:eigenvalueProblem}
	\lambda
	\vec{y}=
	i
	\begin{pmatrix} 
		0 & \mathds{1} \\
		- D & \Gamma 
	\end{pmatrix}
	\vec{y}
	\,,\qquad
	\vec{y}=
	\begin{pmatrix} 
		\vec{x}(0) \\ \dot{\vec{x}}(0)
	\end{pmatrix},
\end{equation}
where we gathered the indices $i$ in a vector notation $\vec{x}(t)$. Energy conservation requires all eigenvalues $\lambda$ to be real, but the ansatz renders the problem complex. However, a suitable superposition of complex eigensolutions always allows to create real solutions with $\vec x(t)\in\mathds{R}^N$.

While equation~\ref{eq:eigenvalueProblem} contains all the information about the eigensolutions, for the topological classification it is advantageous to transform it into a hermitian form. To this end, we apply the transformation
\begin{equation}
	T=
	\begin{pmatrix}
		\sqrt{D} & 0 \\
		0 & i
	\end{pmatrix}
\end{equation}
to $\vec{y} $.
The square root of the matrix $D$ is defined through its spectral decomposition, where the positive branch of the square root of the eigenvalues is chosen. With this we arrive at
\begin{equation}\label{eq:hermitianForm}
	i
	\frac{d}{d t} \vec{\psi}(t)
	=
	H \vec{\psi} (t) \,,\quad \vec{\psi}(t)=e^{-i\lambda t} T\vec{y} \,,\quad 
	 H=
\begin{pmatrix} 
    0  & \sqrt{ D} \\
    \sqrt{ D} & i \Gamma 
\end{pmatrix}.
\end{equation}
As $D$ is symmetric positive-definite and $\Gamma$ is skew-symmetric, the matrix $H$ is hermitian and the differential equation for $\vec{\psi } $ has the sought after form of a Schr\"odinger equation.

The formulation in equation~\ref{eq:hermitianForm} is reminiscent of a single-particle tight-binding problem in quantum mechanics. Therefore, the discussion of topological properties of the eigenvectors $\vec{\psi}$ can be directly carried over. Remember that the topological classification is based on the spatial dimensionality of the problem as well as properties of special local symmetries alone. In particular, the topological properties do not rely on translational symmetries. However, their discussion and definitions are most conveniently introduced for translationally symmetric systems. In this case, the $D$ and $\Gamma$ matrices are periodic and a spatial Fourier transform block-diagonalizes them (one block for each wave vector $\vec{k}$). It follows that $\sqrt{D}$ becomes block diagonal as well, because it shares its eigenvectors with $D$. Hence we will discuss families  $H(\vec{k})$ of Hamiltonians of the form \ref{eq:hermitianForm}. 

Before turning our attention to the topological classification, we comment on two more points: (i) the influence of damping and (ii) the possibility for alternative hermitian forms. Every real system is prone to damping, which in turn affects the eigensolutions in two ways. The eigenvalues acquire an imaginary part and the form of the eigenvectors may change. While a slight change of the eigenvalues does not influence the subsequent discussion, the difference in the eigenvectors may alter the results. Whether or not it obstructs the use or observation of a given topological effect depends on the details of the specific system.

Now to the second point. The transformation leading to equation~\ref{eq:hermitianForm} is not the only way to introduce a hermitian problem. In fact, any decomposition of the form $D=QQ^T$, $Q\in\mathds{R}^{N\times M}$ will allow us to achieve this goal. By introducing the auxiliary variables $\vec{\eta}(t)=Q^T \vec{x}(t)$, we may express equation~\ref{eq:eom} as 
\begin{equation}
	i
	\frac{d}{dt} \begin{pmatrix}
		\vec{\eta} (t) \\ i \dot{\vec{x}}(t) 
	\end{pmatrix}
	=
	\begin{pmatrix}
		0 & Q^T \\
		Q  & i \Gamma
	\end{pmatrix}
	\begin{pmatrix}
			\vec{\eta} (t) \\ i\dot{\vec{x}}(t) 
	\end{pmatrix}.
\end{equation}
The particular choice $Q=\sqrt{D}=Q^T$ has the advantage that (i) $\sqrt{D}$ has the same eigenvectors as $D$, (ii) it uses only as many auxiliary degrees of freedom as needed, (iii) it allows to directly block-diagonalize the problem in absence of $\Gamma$, and (iv) it offers a canonical way how to choose $Q$.

Nevertheless, this is not the only useful choice of $Q$. The starting point for our choice was a given $D$ and $\Gamma$, originating from an effective model. In certain cases however, there is a natural choice of $Q$ along with a physical meaning. Such cases have been considered by Kane and Lubensky \cite{Kane13,Lubensky15}. In their setup the matrix $Q$ corresponds to the equilibrium matrix of a mass-spring model, where $Q$ relates spring tensions to displacements of the attached masses. This allows for a beautiful discussion of (topological) states of self stress in isostatic lattices \cite{Kane13,Lubensky15}. While such states of self stress elude our description, the formulation of Kane and Lubensky is only applicable to the restricted set of isostatic models, which makes it not the favourite choice for the purpose of our discussion. 

\section*{Symmetries}

\begin{figure}[t!]
\centerline{\includegraphics{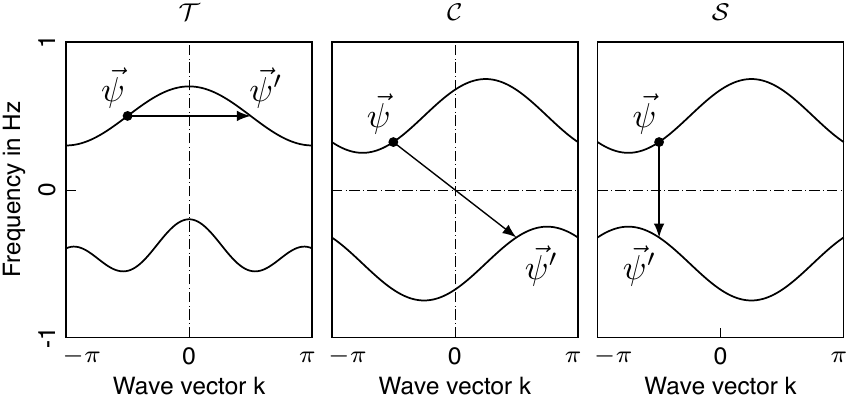}}
\caption{Visualization of $\mathcal{T}$-, $\mathcal{C}$- and $\mathcal{S}$-symmetry by three prototypical band-structures. The presence of a symmetry implies a certain symmetry in the band-structure (but not the other way around), see text.
\label{fig:symmetries}}
\end{figure}
As mentioned before, the classification of electronic systems is based on three symmetries: time-reversal symmetry $\mathcal{T} $, particle-hole symmetry $\mathcal{C}$, and chiral symmetry $\mathcal{S}$. In the quantum mechanical case, these symmetries are represented by (anti-)unitary operators on the single-particle Hilbert space. For the present context of classical mechanical systems, it is important to note that these symmetries are merely a set of constraints on the Bloch Hamiltonians $H(\vec{k})$ \cite{Ryu10}. We state here the form of these constraints and discuss their relation to natural symmetries of mechanical systems below.

We call a system $\mathcal{T}$-symmetric if
\begin{equation}\label{eq:timeReversalSymmetry}
	U_{\mathcal{T} }H(\vec{k})-H(-\vec{k})U_{\mathcal{T} }=0\,,\qquad {U_{\mathcal{T} }}^2= \pm \mathds{1}  \,,
\end{equation}
for some anti-unitary $U_{\mathcal{T} }$ which represents $\mathcal{T} $. For the particle-hole symmetry $\mathcal{C}$, the respective criterion is
\begin{equation}\label{eq:particleHoleSymmetry}
	U_{\mathcal{C} }H(\vec{k})+H(-\vec{k})U_{\mathcal{C} }=0\,,\qquad {U_{\mathcal{C} }}^2= \pm \mathds{1}  \,,
\end{equation}
with $U_{\mathcal{C} }$ anti-unitary. Finally, the chiral symmetry $\mathcal{S}$ we demand
\begin{equation}\label{eq:chiralSymmetry}
	U_{\mathcal{S} }H(\vec{k})+H(\vec{k})U_{\mathcal{S} }=0\,,\qquad {U_{\mathcal{S} }}^2= \mathds{1}  \,,
\end{equation}
for a unitary $U_{\mathcal{S} }$, cf. Fig.~\ref{fig:symmetries}.

For a generic Hilbert space $\mathcal{H} $ there are no additional restrictions on the representations $U$, but here the ``Hilbert space'' has additional structure. Any eigenvector $\vec{\psi }$ is of the form $(\sqrt{D} \vec{x}(0) , \lambda \vec{x}(0))$. Hence, after fixing the first half of the entries of $\vec{\psi} $ the remaining half is known as well. It follows that any (anti-)unitary mapping $U: \mathcal{H} \to \mathcal{H}$ can be written as
\begin{equation}\label{eq:formOfSymmetry} 
	U
	=
	\begin{pmatrix}
		W & 0  \\
		0 & V
	\end{pmatrix},
\end{equation} 
with $V$ and $W$ (anti-)unitary. Let us have a closer look at the three symmetries \ref{eq:timeReversalSymmetry}-\ref{eq:chiralSymmetry} within this framework.

From the definitions, it follows that $H$ has $\mathcal{T}$-symmetry if and only if we can find $V$, such that
\begin{equation}
	V\Gamma(\vec{k})+\Gamma(-\vec{k}) V=0\,,\qquad V D(\vec{k})- D(-\vec{k})V =0\,,
\end{equation}
with $W= V$. We refer to it as $\mathcal{T}$-symmetry, instead of ``time-reversal'', because in the setting of classical mechanics it does no longer correspond to the reversal of time. In case that $\Gamma=0$, there is a generic $\mathcal{T}$-symmetry
\begin{equation}
	U_\mathcal{T} 
	=
	\begin{pmatrix}
		\mathds{1} & 0 \\
		0 & \mathds{1} 
	\end{pmatrix}
	\kappa\,,\qquad {U_\mathcal{T}}^2=1\,,
\end{equation}
where $\kappa$ is the complex conjugation operator. Note that even though $\Gamma$ has the potential to break $\mathcal{T}$-symmetry, $\Gamma\neq 0$ does not imply the absence of it.

For $\mathcal{C}$-symmetry, the conditions to be satisfied are
\begin{equation}
	V\Gamma(\vec{k})-\Gamma(-\vec{k}) V=0\,,\qquad V D(\vec{k})- D(-\vec{k})V =0\,,
\end{equation}
with $W= -V$.
Therefore, for any $D$ and $\Gamma$ we can find a particle hole symmetry
\begin{equation}\label{eq:particleHoleSym}
	U_\mathcal{C} 
	=
	\begin{pmatrix}
		\mathds{1} & 0 \\
		0 & -\mathds{1} 
	\end{pmatrix}
	\kappa\,,\qquad {U_\mathcal{C}}^2=1\,.
\end{equation}
The existence of this omnipresent particle-hole symmetry is nothing but the statement that for every eigensolution, its complex conjugate is also an eigensolution. Its presence is based on $D$ and $\Gamma$ being real. 

In case we have $\mathcal{T}$- and particle-hole symmetry, we can combine the two to obtain a unitary operator $U_{\mathcal{S}}=U_{\mathcal{C}}U_{\mathcal{T}}$. This unitary operator represents a chiral symmetry. We can therefore conclude, that if $\Gamma=0$ we always have a chiral symmetry
\begin{equation}
	U_{\mathcal{S}}
	=
	\begin{pmatrix}
		\mathds{1} & 0 \\
		0 & -\mathds{1} 
	\end{pmatrix}.
\end{equation}
This symmetry is nothing but {\it classical} time-reversal symmetry, as every eigenvector is mapped to itself and the corresponding eigenvalue becomes minus itself.

So far, particle-hole and chiral symmetries were defined with respect to $\omega=0$, meaning that an eigensolution $(\vec{\psi}, \lambda)$ is related to an eigensolution $(\vec{\psi}'=U\vec{\psi}, \omega- \lambda)$ with $\omega=0$, cf. Fig.~\ref{fig:symmetries}. However, for the purpose of topological indices, we can weaken this requirement. A potentially $\vec{k}$-dependent shift in $\omega$ does not change the form of the eigenvectors. Hence, it is sufficient to require the right-hand side of equations~\ref{eq:particleHoleSymmetry} and \ref{eq:chiralSymmetry} to equal to $ 2 \omega(\vec{k}) U_{\mathcal{C}/\mathcal{S}}$ instead of zero. Furthermore, particle-hole and chiral symmetries can also exist only on parts of the band-structure. Which means that it is possible to have this symmetries on a subspace of all the solutions only.

These two generalizations of $\mathcal{C}$- and $\mathcal{S}$-symmetries arise naturally in the setting of mass-spring models.\footnote{Note, that these generalization are not the only ones possible. However, they emerge naturally in our present discussion.} Assume that $D$, which is a real, symmetric matrix and therefore hermitian, has a particle-hole symmetry with respect to $\omega\neq 0$, and that $\Gamma=0$. Then, all the eigenvectors of $H$ with positive eigenvalue have a particle hole symmetry with respect to some $\omega(\vec{k})$, while all the eigenvectors of $H$ with negative eigenvalue have one with respect to $-\omega(\vec{k})$. The matrix $H(\vec{k})$ can be made block-diagonal with the two blocks $\pm \sqrt{D}$ and each corresponding subspace of solutions has a particle-hole (and chiral) symmetry with respect to $\pm \omega(\vec{k})$.

After discussing the above symmetries, we have all the elements we need to establish a topological classification of generic mechanical systems.

\section*{Classification}

\begin{table}[t]
	\centering
	\caption{The tenfold way. The color code is explained in the main text. This table also applies to the high-frequency problem of non-reciprocal metamaterials.}
	\label{tab:tenfoldWay}
	\begin{tabular}{@{\vrule height 9pt depth4pt  width0pt}@{\extracolsep{\fill}}c|ccc|cccc}
		 & \multicolumn{3}{c}{Symmetries} & \multicolumn{3}{c}{Dimensions} \\
		Class & $\mathcal{T}$ & $\mathcal{C}$ & $\mathcal{S}$ & 1 & 2 & 3 \\ \hline
		A    & 0   & 0   & 0 & 0                            &{ \color{blue}$\mathds{Z}$} & 0 \\
		AIII & 0   & 0   & 1 & {\color{red} $\mathds{Z}$ }    & 0 & {\color{red} $\mathds{Z}$} \\\hline
		AI   & $+$ & 0   & 0 & 0 & 0 & 0 \\
		BDI  & $+$ & $+$ & 1 & {\color{red} $\mathds{Z}$} & 0 & 0 \\
		D    & 0   & $+$ & 0 & {\color{myBlue}$\mathds{Z}_2$} & {$\color{blue} \mathds{Z}$} & 0 \\
		DIII & $-$ & $+$ & 1 & {$\color{orange}\mathds{Z}_2$ }&{$\color{orange}\mathds{Z}_2$} & {\color{red}$\mathds{Z}$} \\
		AII  & $-$ & 0   & 0 & 0 & {\color{myBlue}$\mathds{Z}_2$} & {\color{myBlue}$\mathds{Z}_2$} \\
		CII  & $-$ & $-$ & 1 & {$\color{red} 2\mathds{Z}$} & 0 & {\color{orange}$\mathds{Z}_2$} \\
		C    & 0   & $-$ & 0 & 0 & {\color{blue} $2\mathds{Z}$} & 0 \\
		CI   & $+$ & $-$ & 1 & 0 & 0                          & {\color{red}$2\mathds{Z}$} \\\hline
	\end{tabular}
\end{table}

With the mapping of the equations of motion to a hermitian eigenvalue problem we can in principle directly use the classification scheme of non-interacting electron systems \cite{Kitaev09, Ryu10}. However, the specific properties of the local symmetries discussed above warrant a more careful discussion. To make further progress, we highlight the most important concepts behind the electronic classification. For a more detailed review we refer the reader to the excellent recent review by Chiu et al. \cite{Chiu15}. A reader not interested in the details of the derivation might jump straight to tables \ref{tab:tenfoldWay}-\ref{tab:lowFreqGamma} for a reference of possible topological phonon systems and the example section for an illustration of theses tables.

For non-interacting electrons, the ground-state is given by a Slater determinant of all states below the chemical potential. The topological properties are then encoded in the projector $P(\vec{k})$ onto the filled bands. Moreover, one can simplify the discussion by introducing a ``flattened Hamiltonian'' $Q(\vec{k})=\mathds 1-2P(\vec{k})$ which assumes the eigenvalues $\pm 1$ for filled (empty) bands \cite{Chiu15}.

The topological indices are now encoded in the mappings from the Brillouin-zone to an appropriate target space induced by $Q(\vec{k})$. In the absence of any symmetries the target space are the set of complex Grassmanians. In {\em even dimensions}, these mappings are characterized by Chern numbers that lie in $\mathds Z$ (marked in blue in Tab.~\ref{tab:tenfoldWay}). In case that the chiral symmetry $\mathcal S$ is present, the $Q(\vec{k})$ matrices have additional structure. This structure can be used to block-off-diagonlize them \cite{Ryu10, Chiu15}
\begin{equation}
	Q({\vec{k}})=
	\begin{pmatrix}
		0 & q({\vec{k}}) \\
		q^\dag(\vec{k}) & 0 
	\end{pmatrix},
\end{equation}
and to obtain a mapping from the Brillouin zone to the space of unitary matrices. In {\em odd dimensions} the homotopy group of these maps is described by a winding number $\in \mathds Z$ (marked in red in Tab.~\ref{tab:tenfoldWay}). These two types of indices are called the {\em primary indices}.

Additional indices can be derived from the primary ones when more symmetries are present. By constructing families of $d-1$ and $d-2$-dimensional systems whose interpolation constitute a $d$-dimensional Hamiltonian with a primary index, one can establish topologically distinct families of such lower dimensional band-structures through {\em descendant indices}. They are marked in light-blue (light-red) for descendents of the Chern (winding) numbers. Moreover, certain symmetries restrict the primary indices to even values denoted by $2\mathds Z$ in Tab.~\ref{tab:tenfoldWay}. Concrete formulas for the Chern and winding numbers are given in the %
\ifpnas SI. %
\else Appendix. %
\fi%
For general formulas for the descendent indices we refer to \cite{Chiu15} and references therein. This overview concludes our discussion of the electronic classification which is summarized in Tab.~\ref{tab:tenfoldWay}. 

For mechanical systems a few characteristics deserve special attention. First, in a mechanical system, no Pauli principle is available to give a band as a whole a thermodynamic relevance. However, it is clear that the projector to a given number of isolated bands encodes the topological properties of the (high-frequency)\footnote{``High'' is to be understood as ``non-zero''.} gap above these bands. For engineering applications in the respective frequency range this is good enough. Second, before we apply the topological classification of Tab.~\ref{tab:tenfoldWay} blindly to a generic mechanical system it is beneficial to first structure the problems at hand by ``non-topological'' considerations. 

There are two natural properties which divide the mechanical problems into four different classes: (i) A mechanical system can either be made from ``passive'' building blocks, or it can incorporate non-reciprocal elements. In our formulation they distinguish themselves by the absence or presence of a $\Gamma$-term in the Hamiltonian \ref{eq:hermitianForm}. (ii) The formulation of topological indices is rather different for the case where we target the gap around $\omega=0$ (relevant for thermodynamic or ground state properties) or a gap at finite frequencies. In the following we discuss the different combinations of finite versus zero-frequency and reciprocal versus non-reciprocal materials separately. 

{\bf High-frequency non-reciprocal metamaterials.} The presence of $\Gamma \neq 0$ puts the high-frequency problem of non-reciprocal metamaterials on equal footing with the electronic problem. Therefore, no further constraints are imposed and the full Tab.~\ref{tab:tenfoldWay} is explorable. 

\begin{table}[t]
	\centering
	\caption{Indices for high-frequency reciprocal metamaterials with $\Gamma=0$. There is always a $\mathcal T$-symmetry squaring to $+\mathds 1$, which can be augmented to $\mathcal T^*$ squaring to $-\mathds 1$.}
	\label{tab:highFreqGamma0}
	\begin{tabular}{@{\vrule height 9pt depth4pt  width0pt}@{\extracolsep{\fill}}c|ccc|cccc}
		 & \multicolumn{3}{c}{Symmetries} & \multicolumn{3}{c}{Dimensions} \\
		Class & $\mathcal{T}/\mathcal{T}^*$  & $\mathcal{C}$  & $\mathcal{S}$  & 1 & 2 & 3 \\ \hline
		BDI  & $+$ & $+$ & 1 & \color{red}$\mathds{Z}$ & 0 & 0 \\
		DIII & $+/-$ & $+$ & 1 & \color{orange}$\mathds{Z}_2$ &\color{orange} $\mathds{Z}_2$ & 0 \\
		AII  & $+/-$ & 0   & 0 & 0 & \color{myBlue}$\mathds{Z}_2$ & \color{myBlue} $\mathds{Z}_2$ \\
		CII  & $+/-$ & $-$ & 1 & 0 & 0 &\color{orange} $\mathds{Z}_2$ \\
		CI   & $+$ & $-$ & $1$ & $0$ & $0$ & $\color{red}2\mathds{Z}$ \\\hline
	\end{tabular}
\end{table}

{\bf High-frequency reciprocal metamaterials}. For reciprocal high-frequency problems, one can in principle apply the classification scheme to $D$ rather than $H$, as already $D$ is a hermitian matrix.\footnote{Remeber that $\sqrt{D}$ and $D$ share the same eigenvectors and we continue with $H$ to keep the discussion unified.} The reality of $D$ ensures the presence of a $\mathcal T$-symmetry that squares to $+\mathds 1$. One can augment this $\mathcal T$ symmetry to an anti-unitary symmetry $\mathcal T^*$ that squares to $-\mathds 1$ via an appropriate unitary symmetry $U_{\scriptscriptstyle\rm aug}$
\begin{equation}
U_{\mathcal T^*} = U_{\scriptscriptstyle\rm aug} \circ U_{\mathcal T}\quad \mbox{with}\quad U_{\mathcal T^*}^2=-\mathds 1\,.
\end{equation}
However, it is important to note that the simultaneous presence of both a $\mathcal T$- and $\mathcal T^*$-symmetry will force certain indices to vanish. A careful but straight-forward analysis [cf. %
\ifpnas SI%
\else App.%
\fi%
] of the indices results in  Tab.~\ref{tab:highFreqGamma0} relevant for reciprocal high-frequency problems.

\begin{table}[t]
	\centering
	\caption{Indices for low-frequency reciprocal metamaterials with $\Gamma=0$. Both the $\mathcal C$ and $\mathcal T$ symmetry need to be augmented to reach classes where these symmetries square to $-\mathds 1$.}
	\label{tab:lowFreqGamma0}
	\begin{tabular}{@{\vrule height 9pt depth4pt  width0pt}@{\extracolsep{\fill}}c|ccc|cccc}
		 & \multicolumn{3}{c}{Symmetries} & \multicolumn{3}{c}{Dimensions} \\
		Class & $\mathcal{T}/\mathcal{T}^* $  & $\mathcal{C}/\mathcal{C}^*$  & $\mathcal{S}$ & 1 & 2 & 3 \\ \hline
		BDI  & $+$ & $+$ & 1 & $\color{red}\mathds{Z}$ & 0 & 0 \\
		DIII & $+/-$ & $+$ & 1 & $\color{orange} \mathds{Z}_2$ & {\color{orange} $\mathds{Z}_2$ }& 0 \\
		CII  & $+/-$ & $+/-$ & 1 & 0 & 0 &{\color{orange} $\mathds{Z}_2$ }\\ \hline
	\end{tabular}
\end{table}
\begin{table}[t]
	\centering
	\caption{Indices for low-frequency non-reciprocal metamaterials with $\Gamma\neq 0$. Here, only the $\mathcal C$ symmetry needs to be augmented as no generic $\mathcal T$ symmetry is present.}
	\label{tab:lowFreqGamma}
	\begin{tabular}{@{\vrule height 9pt depth4pt  width0pt}@{\extracolsep{\fill}}c|ccc|cccc}
		 & \multicolumn{3}{c}{Symmetries} & \multicolumn{3}{c}{Dimensions} \\
		Class & $\mathcal{T} $  & $\mathcal{C}/\mathcal{C}^*$ & $\mathcal{S}$ & 1 & 2 & 3 \\ \hline
		BDI  & $+$ & $+$ & 1 & $\color{red}\mathds{Z}$ & 0 & 0 \\
		D    & 0   & $+$ & 0 & $\color{myBlue}\mathds{Z}_2$ & $\color{blue}\mathds{Z}$ & 0 \\
		DIII & $-$ & $+$ & 1 & $\color{orange}\mathds{Z}_2$ & $\color{orange}\mathds{Z}_2$ & $\color{red}\mathds{Z}$ \\
		CII  & $-$ & $+/-$ & 1 & 0 & 0 & $\color{orange}\mathds{Z}_2$ \\
		C    & 0   & $+/-$ & 0 & 0 & $\color{blue}2\mathds{Z}$ & 0 \\ \hline
	\end{tabular}
\end{table}

{\bf Low-frequency reciprocal metamaterials}. Topological band-structures with non-trivial gaps around zero frequency are relevant for floppy modes in static problems \cite{Paulose15} or thermodynamic properties \cite{Sussman15} of jammed granular media. As argued above, the structure of the equations of motion imply a $\mathcal C$-symmetry around $\omega=0$. In the absence of $\Gamma$, an additional $\mathcal T$ symmetry is present. Both this built-in symmetries canonically square to $+\mathds 1$. As in the case of high-frequency reciprocal materials, one can augment these symmetries by unitary symmetries to reach classes where the augmented ones square to $-\mathds 1$. Tab.~\ref{tab:lowFreqGamma0} summarizes the resulting possibilities for topological indices in this setup.

{\bf Low-frequency non-reciprocal metamaterials}. Similarly to the high-frequency non-reciprocal metamaterials, the generic $\mathcal T$-symmetry is absent here. Hence, there can arise effective $\mathcal T$ symmetries that either square to $+ \mathds 1$ or $-\mathds 1$ without the need to augment the generically present one in order to reach classes where $\mathcal T^2=-\mathds 1$. Given that we deal with the gap at $\omega=0$, however, guarantees the generic $\mathcal C$-symmetry which in turn can be enriched to one that squares to $-\mathds 1$. The resulting possible topologies are shown in Tab.~\ref{tab:lowFreqGamma}. 

For the case of zero-frequency indices, the construction of $H$ with the help of $\sqrt{D}$ necessary leads to trivial phases, cf. %
\ifpnas SI. %
\else
App. %
\fi%
 However, in Refs. \cite{Kane13, Lubensky15} it was shown how a decomposition $D=QQ^T$ allowing for non-trivial $\mathds Z$-indices in class BDI can be constructed for Maxwell frames. How one can construct similar formulations for the other symmetry classes shown in Tabs.~\ref{tab:lowFreqGamma0} and \ref{tab:lowFreqGamma} is an interesting open problem.

\section*{Examples}
To clarify and reinforce our approach we provide a set of examples. We directly consider discrete models. An example on how to extract a discrete description of a continuum model is provided in the %
\ifpnas SI. %
\else
Appendix. %
\fi%
The degrees of freedom are assumed to be ideal one or two dimensional pendula. The desired $D$-matrix can be obtained by coupling the different pendula by springs. To encode negative coupling elements, or in case of geometrical obstructions, it might be required to replace a spring coupling by a more involved coupling composed of springs and deflection levers, as e.g. in Ref.~\cite{Susstrunk15}. Note, that while we consider pendula as our local oscillator, all examples are generic and can be applied to any set of mechanical modes.

\begin{figure}[t]
	\centerline{\includegraphics[width=3.5cm]{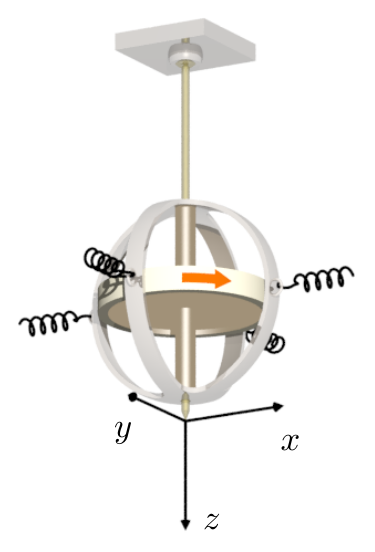}}
	\caption{Coordinate system for a spinning gyroscope.
    \label{fig:gyro}}
\end{figure}

The last ingredient we need is a $\Gamma\neq 0$. One option is to engage the Lorentz force, which directly provides such a coupling. Another possibility is to use spinning tops, or gyroscopes as in Refs.~\cite{Nash15, Wang15a}. We consider a symmetric gyroscope with a fixed point (different from the center of mass) about which it can rotate, cf. Fig.~\ref{fig:gyro}. For our considerations, there will be no external moment along the principal axis passing through the center of mass, rendering this rotation a conserved quantity. Hence, there are only two degrees of freedom left.

In a constant gravitational field, we can use the direction of the field to define a $z$-axis. The potential energy of the gyroscope has a minimum and one can linearize the problem about this minimum. The resulting problem has two effective degrees of freedom which we choose to be displacements along the $x$ and $y$ direction. The equation of motion for the linearized system is then of the form, cf. %
\ifpnas SI,%
\else
 App.,%
\fi
\begin{equation*}
	\begin{pmatrix}
		\dot{x} \\ \dot{y} \\
		\ddot {x} \\ \ddot{y}
	\end{pmatrix}
	=
	\begin{pmatrix}
		0 & 0 & 1 & 0  \\
		0 & 0 & 0 & 1  \\
		-\mu & 0 & 0 & \gamma \\
		0 & -\mu & -\gamma & 0
	\end{pmatrix}
	\begin{pmatrix}
		x \\ y \\
		\dot {x} \\ \dot{y}
	\end{pmatrix}
	+
	\begin{pmatrix}
		0 \\ 0 \\
		M_x \\ M_y
	\end{pmatrix},
\end{equation*}
where $\gamma$ is proportional to the spinning speed of the gyroscope and $M_{x/y}$ are external moments coupling to it. Such moments arise, e.g., from the couplings to  neighboring degrees of freedom. For multiple gyroscopes, this allows us to obtain
\begin{equation}\label{eq:gammaMatrix}
	\Gamma
	=
	\gamma
	\begin{pmatrix}
		0 & \mathds{1} \\
		- \mathds{1} & 0 
	\end{pmatrix}.
\end{equation}

These are all the elements we need to discuss the following examples. While every model has a high frequency and a low frequency symmetry part, we are only looking at the former, where the generic particle-hole symmetry is irrelevant. For a detailed discussion of certain low-frequency models we refer to Refs.~\cite{Kane13,Lubensky15}.

We start with the simplest possible one dimensional model with a non-trivial index below. After its discussion, we show how one can combine several copies of such simple building blocks to reach a number of other symmetry classes in one dimension. Finally, we provide each an example of a two-dimensional system with a non-vanishing Chern number and a model where we employ the idea of symmetry-enrichment. 

\subsection*{Class BDI in 1D}
Probably the simplest model available is the analogue of the Su-Schrieffer-Heeger (SSH) model \cite{Su79}. It can be realized through a chain of one dimensional pendula, coupled through springs with alternating spring constants $t$ and $t'$. Its dynamics is governed by
\begin{equation}\label{eq:BDI1D}
	D(k;t,t')
	=
	\begin{pmatrix}
		\mu & -t-t' e^{-ik} \\
		-t-t' e^{ik} & \mu
	\end{pmatrix},\qquad
	\Gamma=0\,,
\end{equation}
and $\mu>|t|+|t'|$ for positive definitness.

The model has a $\mathcal S$-symmetry (chiral symmetry), which can already be seen on the level of the $D$ matrix:
\begin{equation*}
	U_{\mathcal{S}}
	D(k)+D(k)U_{\mathcal{S}}
	=2\mu U_{\mathcal{S}} \,,\qquad 
	U_\mathcal{S}=
	\begin{pmatrix}
		1 & 0 \\
		0 & -1
	\end{pmatrix}.
\end{equation*}
The symmetry translates into a $\mathcal S$-symmetry of $\pm \sqrt{D} $, which are the two blocks of $H(\vec{k} )$ after block-diagonalizing it:
\begin{equation}\label{eq:blochDiagonalizing}
	T
	H(k)
	T^\dagger
	=
	\begin{pmatrix}
		\sqrt{D}(k )  & 0 \\
		0 & -\sqrt{D}(k )
	\end{pmatrix},\quad T=\frac{1}{\sqrt{2} } 
	\begin{pmatrix}
		\mathds{1}  & \mathds{1} \\
		\mathds{1} & -\mathds{1}
	\end{pmatrix}.
\end{equation}

In addition, the model has $\mathcal{T}$-symmetry and therefore $\mathcal C$-symmetry as well, which puts it into symmetry class BDI. This class features a winding number through its $Q(k)$ matrix
\begin{equation}
	Q(k)
	=
	\begin{pmatrix}
		0 & q(k) \\
		q^*(k) & 0
	\end{pmatrix},\qquad q(k)=c(k)(t+t'e^{-ik})\,
\end{equation}  
with $c(k)\in \mathds R$. The matrix is already in block-off-diagonal form and hence, the winding number $n\in\mathds{Z}$ is given by, cf. %
\ifpnas SI,%
\else
 App.,%
\fi
\begin{equation*}
	n=
	\frac{i}{2\pi}\int q(k)^{-1} q'(k) dk= \begin{cases}
		0 & t<t' \\
		-1 & t>t' \\
	\end{cases}\,.
\end{equation*}

The band-structure of the periodic system is shown in the left part of Fig.~\ref{fig:BDItoAIII}, and the eigenfrequencies of the open system are given in the middle part of the figure at the point $\gamma=0$ (see below). Up to here, we were free to discuss the problem in terms of $D(k)$ instead of $\sqrt{D}(k)$. However, this is no longer possible once $\Gamma\neq 0$, as considered in the next example.

\subsection*{Class AIII in 1D}
\begin{figure}[t]
\centerline{\includegraphics{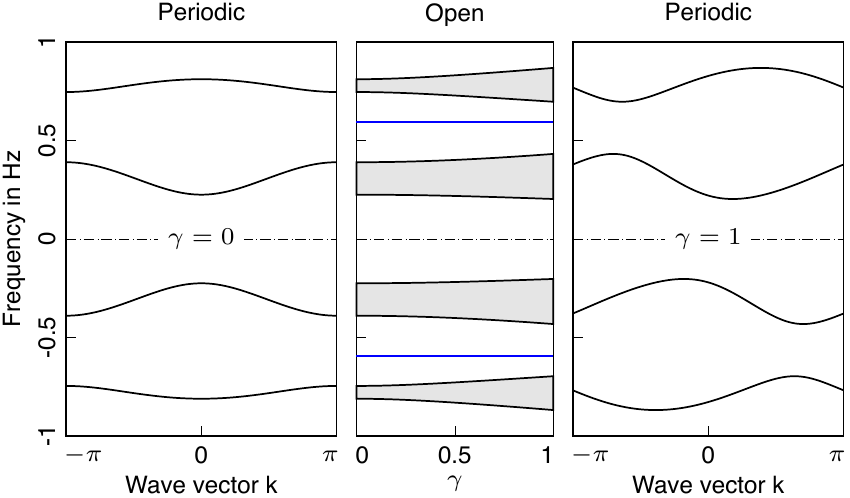}}
\caption{Spectra of examples belonging to classes BDI ($\gamma=0$) and AIII ($\gamma\neq 0$). Left: The band-structure of the BDI model given in equation~\ref{eq:BDI1D}. Middle: Spectrum of an open AIII chain as a function of $\gamma$. Blue lines denote edge modes, whereas the gray areas represent the bulk modes. Right: The bandstructure of the AIII model for $\gamma=1$. Parameters chosen to obtain the figures are: $t=2\,\text{Hz}^{2}$, $t'=10\,\text{Hz}^{2}$ and $\mu=14\,\text{Hz}^{2}$.\label{fig:BDItoAIII}
}
\end{figure}

The above model is now supplemented by a non-vanishing $\Gamma$ matrix. This breaks the $\mathcal{T}$- and the particle-hole symmetry, but the chiral symmetries on the two subspaces (positive / negative eigenfrequencies) are left invariant. In the case that $|\gamma|\ll |t-t'|$, all spectral gaps remain open and hence the topological index will not change. The evolution of the gap as well as of the edge mode (which stays invariant) for increasing $\gamma$ is shown in the middle part of Figure~\ref{fig:BDItoAIII}. An exemplary band-structure for $\gamma=1$ can be seen in right part of Figure~\ref{fig:BDItoAIII}.

Breaking $\mathcal{T}$- and $\mathcal C$-symmetry of the BDI model did not change the topological index, because the index relies on chiral symmetry only.

\subsection*{Class D in 1D}

To break the $\mathcal S$-symmetry while keeping $\mathcal C$-symmetry we need to add further degrees of freedom. Starting point are two copies, $D(k;t,t')$ and $D(k;s,s')$, of the above BDI model
\begin{equation}
    \label{eq:BDI1DtwoCopies}
	D(k;t,t',s,s')
	=
	\begin{pmatrix}
		D(k;t,t') & 0 \\
		0 & D(k;s,s')
	\end{pmatrix}.
\end{equation}
We assume that both share the same $\mu$. For $\Gamma=0$, the model belongs to BDI and the winding number of the lowest two bands is given by
\begin{equation*}
	n=
	\begin{cases}
		0 & t<t'\text{ and }s<s' \\
		1 & (t>t'\text{ and }s<s')\text{ or } (t<t'\text{ and }s>s')\\
		2 & t>t'\text{ and }s>s'
	\end{cases}.
\end{equation*}

By choosing $t\neq s$ or $s' \neq t'$, and turning on $\Gamma\neq 0$,
we break all the symmetries except for the high-frequency $\mathcal C$-symmetry
\begin{equation*}
	U_\mathcal{C}
	=
	\begin{pmatrix}
		1 & 0 & 0 & 0 \\
		0 & -1 & 0 & 0 \\
		0 & 0 & 1 & 0 \\
		0 & 0 & 0 & -1
	\end{pmatrix}
	\kappa\,.
\end{equation*}
This puts the model into symmetry class D.

\begin{figure}[tbh]
\centerline{\includegraphics{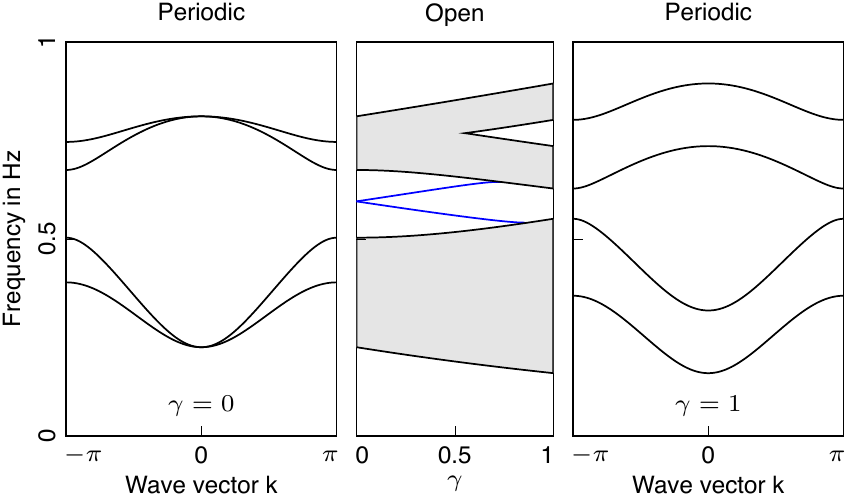}}
\caption{Spectra of examples belonging to classes BDI ($\gamma=0$) and D (trivial) ($\gamma\neq 0$). Left: The band-structure of the periodic BDI model given in equation~\ref{eq:BDI1DtwoCopies}. Middle: Spectrum of the open D model. The parity of the winding number is even, therefore the topological edge modes are not protected upon turning on $\gamma\neq 0$. Right: The bandstructure of the D model for $\gamma=1$. Parameters chosen to obtain the figures are: $t=2\,\text{Hz}^{2}$, $t'=10\,\text{Hz}^{2}$, $s=4\,\text{Hz}^{2}$, $s'=8\,\text{Hz}^{2}$ and $\mu=14\,\text{Hz}^{2}$.
\label{fig:BDItoDtrivial}
}
\end{figure}

\begin{figure}[tbh]
\centerline{\includegraphics{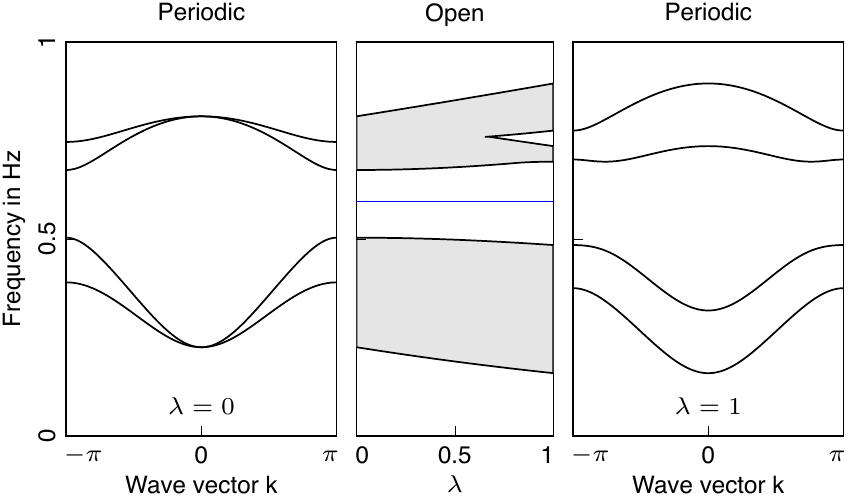}}
\caption{Spectra of examples belonging to classes BDI ($\gamma=0$) and D (non-trivial) ($\gamma\neq 0$). Here, the parity of the winding number is odd and hence the topological edge mode is protected even when $\gamma\neq 0$. Parameters chosen to obtain the figures are: $t=2\,\text{Hz}^{2}$, $t'=10\,\text{Hz}^{2}$, $s=8\,\text{Hz}^{2}$, $s'=4\,\text{Hz}^{2}$ and $\mu=14\,\text{Hz}^{2}$.
\label{fig:BDItoDnonTrivial}
}
\end{figure}

The $\mathds{Z} $ index gets reduced to a $\mathds{Z}_2 $ index,
\begin{equation*}
	p=
	\begin{cases}
		0 & (t-t')(s-s')>0 \\ 
		1 & (t-t')(s-s')<0 
	\end{cases}\,,
\end{equation*}
the parity of the winding number. In the case that $p=0$, the breaking of the $\mathcal S$-symmetry makes the cases $n=0$ and $n=2$ equivalent, as the two edge modes can hybridize and disappear from the gap, see Fig.~\ref{fig:BDItoDtrivial}. In case that $p=1$, the single edge mode from the BDI model remains as displayed in Fig.~\ref{fig:BDItoDnonTrivial}.

\subsection*{Class A in 2D}
The topology of the discussed one-dimensional models relied on the presence of a $\mathcal S$- ($\mathcal C$-) symmetry. The next model we look at does not rely on any symmetries at all and the topological index will be the Chern number. To obtain a non-vanishing Chern number, we need to break $\mathcal{T}$-symmetry by choosing $\Gamma\neq 0$ and therefore we need to have at least two degrees of freedom per unit cell. The $\Gamma$ matrix can only take the form from equation~\ref{eq:gammaMatrix} which leaves us with finding a suitable $D$ matrix.

To this end, it is helpful to transform $H(\vec{k})$, as in equation~\ref{eq:blochDiagonalizing}, to
\begin{equation*}
	TH T^\dagger
	=
	\begin{pmatrix}
		\sqrt{D}+\frac{i}{2} \Gamma & -\frac{i}{2} \Gamma \\
		-\frac{i}{2} \Gamma & -\sqrt{D}+\frac{i}{2} \Gamma 
	\end{pmatrix},
\end{equation*}
and define 
\begin{equation}
	\tilde H(k_x,k_y;\alpha)
	=
	\begin{pmatrix}
		\sqrt{D}+\frac{i}{2} \Gamma & -\alpha\frac{i}{2} \Gamma \\
		-\alpha\frac{i}{2} \Gamma & -\sqrt{D}+\frac{i}{2} \Gamma 
	\end{pmatrix}.
\end{equation}
By varying $\alpha\in [0,1]$, we can continuously deform $H(k_x,k_y)$ into a model with two decoupled blocks $\pm \sqrt{D}+\frac{i}{2} \Gamma$. If the bulk-gaps remain opened during the interpolation from $\alpha=0$ to $\alpha=1$, the Chern number of any subspace will not change, and we can focus on constructing non-trivial subblocks of $\tilde H(k_x,k_y;0)$.

We now focus on the block characterized by $\sqrt{D}+\frac{i}{2} \Gamma$. This matrix is hermitian and can be written as
\begin{equation*}
	\sqrt{D}(k_x,k_y)+\frac{i}{2} \Gamma = \mu(k_x,k_y) \mathds{1} + \vec{d}(k_x,k_y) \cdot \vec{\sigma } \,,
\end{equation*}
where $\sigma_i$ are the Pauli matrices
\begin{equation*}
	\sigma_1
	=
	\begin{pmatrix}
		0 & 1 \\
		1 & 0
	\end{pmatrix},\quad 
	\sigma_2
	=
	\begin{pmatrix}
		0 & -i \\
		i & 0
	\end{pmatrix},\quad
	\sigma_3
	=
	\begin{pmatrix}
		1 & 0 \\
		0 & -1
	\end{pmatrix},
\end{equation*}
and $\vec{d}(k_x,k_y)\in\mathds{R}^3  $ a vector with real coefficients. The $d$ vector contains all the information about the eigensolutions of the problem and therefore also about the Chern numbers of the bands. In case that $|\vec{d} (k_x,k_y)|\neq 0$ for all $k_x$ and $k_y$, we can define $\vec{n}(k_x,k_y)=\vec{d}/|\vec{d} |$. Upon varying $k_x$ and $k_y$ through the Brillouin zone $\vec{n} $ traces out a closed surface in $\mathds{R}^3$. It can be shown, that the number of net encircling of the origin by this surfaces corresponds to the Chern number of the lower band \cite{Bernevig13}. 

A possible choice of coefficients giving rise to a non-trivial band-structure is
\begin{equation}\label{eq:qmDvector}
	\begin{aligned}
		d_1 &=  \cos{k_x}\,, \\
		d_2 &=  \sin{k_x}+\sin{k_y}-\gamma\,, \\
		d_3 &=  \cos{k_y}  \,.
	\end{aligned}
\end{equation}
Owing to the fact that $\sqrt{D}$ and $D$ share the same eigenvalues, it is easy to see that the dynamical matrix is parameterized by
\begin{equation}
	D(k_x,k_y) = \tilde{\mu} \mathds{1} +t\, \vec{d}(k_x,k_y;\gamma=0) \cdot \vec{\sigma } \,,
\end{equation}
for some suitable $\tilde{\mu}$ and $t$.

\begin{figure}[tbh]
\centerline{\includegraphics{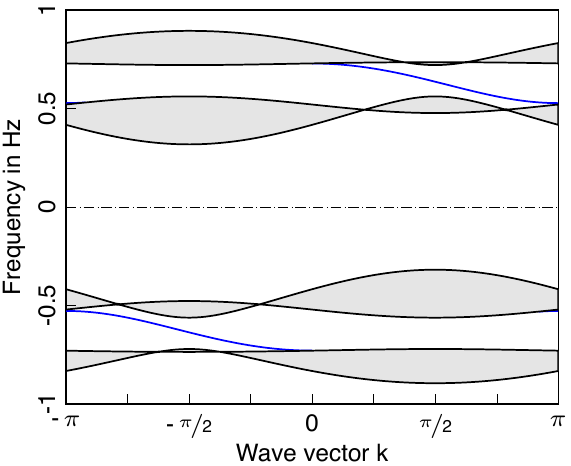}}
\caption{%
Spectrum of the class A model on a semi-infinite cylinder as a function of the wave vector around the cylinder. Gray areas represent a continuum of bulk modes, whereas blue lines denote the chiral surface modes. Parameters chosen to obtain the figure are: $t=5\,\text{Hz}^2$, $\tilde\mu=16\,\text{Hz}^2$ and $\gamma=1\,\text{Hz}^2$.
\label{fig:A}}
\end{figure}

The approximative argument at $\alpha=0$ is supported by a numerical calculation for $\alpha=1$, which confirms the presence of a non-zero Chern number. In addition, we show the spectrum of a semi-infite cylinder in Figure~\ref{fig:A}, revealing the existence of an edge mode within the bulk-gap.

The presented model is a minimal model in the sense of required degrees of freedom. However, it is probably not the simplest model for an actual implementation. For such a purpose, a simpler model can be found in Ref.~\cite{Nash15}.

\subsection*{Class AII in 2D}

Up to here, all examples we looked at where based on symmetries which square to $+\mathds 1$. However, we can also supplement symmetries to obtain new symmetries which can square to $-\mathds 1$. As an example we discuss the quantum spin hall like system presented in Ref.~\cite{Susstrunk15}. It mimics a Hofstadter model \cite{Hofstadter76} at $\Phi=\frac{1}{3}$ flux plus its time-reversed copy. Its dynamical matrix is
\begin{align*}
	D(\vec{k} )
	&=
	\begin{pmatrix}
		D_1(\vec{k} ) & D_2(\vec{k} ) \\
		-D_2(\vec{k} ) & D_1(\vec{k} )
	\end{pmatrix}, \\
	D_1(\vec{k} )
	& =
	-\mu\mathds{1}
	+2t
	\begin{pmatrix}
		2\cos(k_y) & 1 & e^{ik_x} \\
		1 & -\cos(k_y) & 1 \\
		e^{-ik_x} & 1 & -\cos(k_y)
	\end{pmatrix},\\
	D_2(\vec{k} )
	& =
	i 2 \sqrt{3}t \sin(k_y)
	\begin{pmatrix}
		0 & 0 & 0 \\
		0 & 1 & 0 \\
		0 & 0 & -1
	\end{pmatrix},
\end{align*}
and $\Gamma=0$.

\begin{figure}[t]
\centerline{\includegraphics{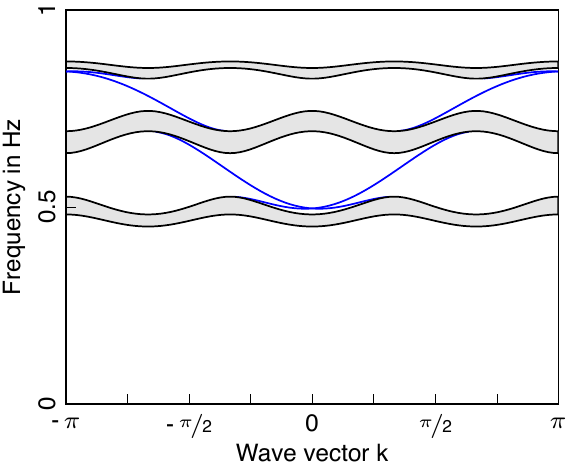}}
\caption{Spectrum of the class AII model on a semi-infinite cylinder as a function of the wave vector around the cylinder. Parameters chosen to obtain the figure are: $t=2\,\text{Hz}^2$ and $\mu=(6+2\sqrt{3})t$. \label{fig:AII}}
\end{figure}

The structure of $D(\vec{k})$ carries a $\mathcal{T}$-symmetry whose anti-unitary representation is just given by the complex conjugation $\kappa$, i.e., $U_{\mathcal T}=\mathds{1} \kappa$. Therefore this symmetry squares to $\mathds 1$. However, there is an additional structure which allows to generate an augmented symmetry $\mathcal T^*$ 
\begin{equation*}
	U_{\mathcal{T^*} }= U_{\scriptscriptstyle\rm aug} \circ \kappa=
	\begin{pmatrix}
		0 & \mathds{1} \\
		-\mathds{1} & 0 
	\end{pmatrix}\kappa\,,\qquad {U_{\mathcal{T^*}}}^2=-\mathds{1}\,,
\end{equation*}
which gets liftet to a $\mathcal{T}^*$ symmetry of $H(\vec{k})$. Otherwise there are no relevant symmetries present away from $\omega=0$, which puts the problem into symmetry class AII. Repeating the calculation from the previous model results in Figure~\ref{fig:AII}. For further details on this model we refer directly to Ref.~\cite{Susstrunk15}.

\section*{Conclusions}

In summary, we have developed a framework to map the equations of motion of a set of coupled linear mechanical oscillators to a quantum mechanical tight-binding problem. Using this mapping we showed how one can import the topological classification of non-interacting electron systems to the realm of classical mechanical metamaterials. Using the presence or absence of non-reciprocal elements as a key aspect of metamaterials, we further adapted the electronic classification to mechanical problems.

With our work we provide the stage for the development of potentially new classes of materials, where topological boundary modes can be used to provide a specific functionality. Moreover, we help to clarify the recent literature in the field, where topological phonon modes have been predicted without an overarching framework. We hope, that with the extensive example section we provided the reader with the tools and concepts to construct more topological phonon models using simple building blocks. 

Many new directions in the field of topological mechanical metamaterials are still unexplored. Obvious problems to be solved are the presentation of a topological surface mode in a two or three dimensional continuous material or the miniaturization of the effects observed at the centimeter scale down to micron scale. Moreover, examples of materials in many of the possible symmetry classes characterized in this report have neither been theoretically proposed nor experimentally implemented. We hope that with this work, we stimulate research in this direction. Moreover, our results also provide the framework to import ideas based on crystalline symmetries.

Finally, the efficient characterization of model-materials according to their topological properties is an important open problem. In electronic systems the search for topological band-structures is now routinely done using high-throughput ab-initio calculations in combination with advanced numerical tools \cite{z2pack} to determine the topological indices. Our framework should provide the basis for a similar approach in mechanical metamaterials and therefore open the route to various applications based on topological boundary modes.

\ifpnas

\acknow{We acknowledge fruitful discussions with O.R. Bilal, T. Bzdu\v{s}ek, C. Daraio, and A. Soluyanov. This work was supported by the Swiss National Science Foundation.}

\showacknow

\else

\begin{acknowledgments}
	We acknowledge fruitful discussions with O.R. Bilal, T. Bzdu\v{s}ek, C. Daraio, and A. Soluyanov. This work was supported by the Swiss National Science Foundation.
\end{acknowledgments}

\fi

\ifpnas

\renewcommand{\thefigure}{S\arabic{figure}}
\renewcommand{\theequation}{S\arabic{equation}}

\subsection*{Supporting Information (SI)}

\else
 
\appendix
\renewcommand{\thefigure}{A\arabic{figure}}
\renewcommand{\theequation}{A\arabic{equation}}

\section*{Appendix}

\fi

\subsubsection*{Example on how to obtain topological indices \ifpnas (SI)\fi}
\label{app:definitions}

As promised we provide formulas for the primary indices in one, two, and three dimensions.
The Chern number in two dimension as a functional of $Q$ is given by \cite{Chiu15}
\begin{equation}
	\begin{aligned}
		C &= \frac{-i}{16 \pi} \int_\text{BZ} \tr[Q(dQ)^{2}] \\
		 &= \frac{-i}{16 \pi} \int_\text{BZ}dk^2 \varepsilon^{\mu\nu} \tr[Q(\partial_{k_\mu} Q)(\partial_{k_\nu} Q)]\,.
	\end{aligned}
\end{equation}
The winding number in odd dimensions is given by \cite{Ryu10}
\begin{equation}
		\nu_1 = \frac{i}{2\pi} \int_\text{BZ} dk \tr [q^{-1}\partial_k q] 
\end{equation}
in one dimension and
\begin{equation}
		\nu_3 = \frac{1}{24 \pi^2} \int_\text{BZ} dk^3 \varepsilon^{\mu\nu\rho} \tr[(q^{-1}\partial_{k_\mu} q)(q^{-1}\partial_{k_\nu} q)(q^{-1}\partial_{k,\rho} q)]
\end{equation}
in three dimensions. These two expressions can be unified using the winding number density
\begin{equation}
	\omega_{2n+1}[q]= \frac{i(-i)^n n!}{(2n+1)! (2\pi)^{n+1}}\tr[(q^{-1}dq)^{2n+1}]\,,
\end{equation}
yielding
\begin{equation}
	\nu_{2n+1}[q]=\int_\text{BZ} \omega_{2n+1}[q]\,.
\end{equation}

Using these expressions for the winding number, we now argue how a combined presence of $\mathcal T$- ($\mathcal C$-) symmetries squaring to $+\mathds 1$ and $-\mathds 1$ leads to the vanishing of certain indices. Let us introduce the notation for the ``signature'' of the symmetry class $(\epsilon_{\mathcal T}, \epsilon_{\mathcal C})$, with $\epsilon_{\alpha}=\pm 1$ denoting that the respective symmetries square to $\epsilon_{\alpha}\mathds 1$. Different signatures dictate different constraints on the winding number density. 

Let us start with $\epsilon_{\mathcal T}\epsilon_{\mathcal C}=-1$. Here, we have \cite{Ryu10, Schnyder08}
\begin{equation}
	q^T(- \vec{k})=\epsilon_{\mathcal T} q(\vec{k})\,,\qquad 
	\epsilon_{\mathcal T}
	=\begin{cases}
		-1 & \text{DIII} \\
	    \phantom{-}1 & \text{CI}
	\end{cases}\,,
\end{equation}
giving rise to the constraint
\begin{equation}
	\omega_{2n+1}[q(\vec{k})]=(-1)^{n+1} \omega_{2n+1}^*[q(-\vec{k})]
\end{equation}
for the winding number density. This leads to a vanishing winding number in one dimension. For $\epsilon_{\mathcal T}\epsilon_{\mathcal C}=+1$ one finds \cite{Ryu10, Schnyder08}
\begin{equation}
	U_{\mathcal T} q^*(-\vec{k})U_{\mathcal T}^{-1}= q(\vec{k})\,,
\end{equation}
which in turn leads to 
\begin{equation}
	\omega_{2n+1}[q(\vec{k})]=(-1)^{n} \omega_{2n+1}^*[q(-\vec{k})]\,.
\end{equation}
This results in a vanishing winding number in three dimensions. For cases where we have two signatures, e.g., $(1,1)$ and $(1,-1)$ due to the enrichment of the $\mathcal C$-symmetry by a unitary symmetry to a $\mathcal C^*$-symmetry (see main text), both constraints apply and one loses additional indices.

\subsubsection*{Low-frequency physics in class BDI \ifpnas (SI)\fi}
\label{app:trivial}
Materials in class BDI carry a $\mathds{Z}$ index as topological classification. For the low-frequency physics, this index can, e.g., be relevant for states of self stress \cite{Kane13, Lubensky15}. However, if we choose $Q=\sqrt{D}$ these states are not accessible, and we find that the $\mathds{Z} $ index is always trivial, which can be seen from the following explicit calculation.

To proof this claim we start from $\Gamma=0$. In this case, the $H$-matrix is given by
\begin{equation}
	H(\vec{k})
	=
	\begin{pmatrix}
		0 & \sqrt{D}(\vec{k})  \\
		\sqrt{D}(\vec{k}) & 0
	\end{pmatrix}
\end{equation}
and the eigenfunctions can be found to be
\begin{equation}
	\vec{\psi}_\pm
	=
	\begin{pmatrix}
		\vec{\phi} \\
		\pm \vec{\phi}  
	\end{pmatrix} ,\qquad
	H \vec{\psi }_\pm=\pm \lambda \vec{\psi }_\pm\,, \qquad \sqrt{D} \vec{\phi}  =\lambda \vec{\phi}\,.
\end{equation}
The projector onto the negative bands is therefore given by
\begin{equation}
	P(\vec{k})
	=
	\frac{1}{2} 
	\begin{pmatrix}
		\mathds{1} & -\mathds{1} \\
		-\mathds{1} & \mathds{1}
	\end{pmatrix},
\end{equation}
and the $Q$-matrix is obtained to be 
\begin{equation}
	Q(\vec{k} )=\mathds{1} - 2P( \vec{k})
	=
	\begin{pmatrix}
		0 & \mathds{1} \\
		\mathds{1} & 0
	\end{pmatrix}.
\end{equation}
This $Q$-matrix has a trivial $\mathds{Z} $ index which concludes the proof in case $\Gamma=0$. The general statement for arbitrary $\Gamma$ follows from the fact that $\Gamma$ cannot close the spectral gap at $\omega=0$, and therefore cannot change the $\mathds{Z} $ index, as shown next.

We show below that $\Gamma$ cannot induce a gap-closing at $\omega=0$, i.e., if there is a gap for some value of $\Gamma$, there will be a gap for any $\Gamma$. A gap-closing at $\omega=0$ requires at least one eigensolution of $H$ with an eigenvalue equal zero, or equivalently, we need
\begin{equation}\label{eq:det0}
	\det \begin{pmatrix}
		0 & Q^T  \\
		Q & i\Gamma
	\end{pmatrix}
	=0\,.
\end{equation}
From
\begin{equation}
	\det A \det H=\det AH\,,
\end{equation}
together with
\begin{equation}
	A=\begin{pmatrix}
		0 & \mathds{1}   \\
		\mathds{1}  & 0
	\end{pmatrix}, \qquad |\det A|= 1\,,
\end{equation}
we find
\begin{equation}
	\left|
	\det \begin{pmatrix}
		0 & Q^T  \\
		Q & i\Gamma
	\end{pmatrix}
	\right|
	=
	\left|
	\det
	\begin{pmatrix}
		Q & i\Gamma \\
		0 & Q^T
	\end{pmatrix}
	\right|
	=
	\left|
	\det
	\begin{pmatrix}
		Q & 0 \\
		0 & Q^T
	\end{pmatrix}
	\right|,
\end{equation}
where the second equality follows from Laplace's formula. This shows that the determinant in \ref{eq:det0} is independent of $\Gamma$,
which proofs the statement.

\subsubsection*{Linearized equations of motion of a gyroscope \ifpnas (SI)\fi}
\label{app:gyro}
\begin{figure}[bt]
	\centerline{\includegraphics{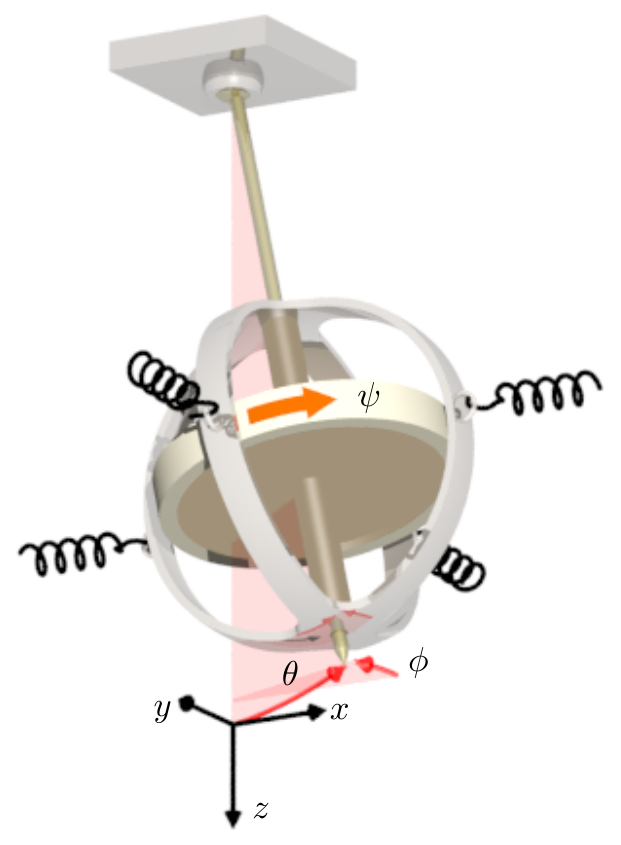}}
	\caption{Coordinate systems for a spinning gyroscope.\label{fig:gyroCoords}}
\end{figure}

We intend to describe the motion of the gyroscope close to its potential minimum, such that the motion of the tip can be described by two cartesian coordinates $x$ and $y$. To give a general expression for the Lagrangian of the gyroscope, we start from a description in Euler angles $\theta$, $\phi$, $\psi$ such that
\begin{equation}
	\begin{aligned}
		x &=R \sin{\theta} \cos{\phi}\,, \\  
		y &=R \sin{\theta} \sin{\phi}\,,
	\end{aligned}
\end{equation}
and $\psi$ describes the rotation with respect to the axis through the center of mass, cf. Figure~\ref{fig:gyroCoords}.

In terms of the principal axis of the gyroscope, its Lagrangian is given by
\begin{equation}
	L=\frac{1}{2} I_1(\Omega_1^2+\Omega_2^2)+\frac{1}{2} I_3 \Omega_3^2-V\,,
\end{equation}
where $V$ denotes potential energy due to gravity and couplings to neighboring gyroscopes. $\Omega_i$ are the angular velocities with respect to the principle axes and $I_i$ the corresponding moments of inertia. The gyroscope is assumed to be symmetric, such that $I_1=I_2$ and $I_3$ is associated with the rotation with respect to the axis through the center of mass. In Euler angles the Lagrangian takes the form
\begin{equation}
	L=\frac{I_1}{2}(\dot{\theta}^2+\dot{\phi}^2 \sin^2\theta)+\frac{I_3}{2} (\dot{\psi}+\dot{\phi} \cos{\theta})^2-V\,.
\end{equation}	
In case that $\frac{\partial V}{\partial \psi}=0 $, which is the case we consider, the Euler-Lagrange equation for $\psi$ reads
\begin{equation}
	\frac{d}{dt} [I_3(\dot{\psi}+\dot{\phi} \cos{\theta})]=0
\end{equation}
and we find the conserved quantity $\Omega=\dot{\psi}+\dot{\phi} \cos{\theta}$.

The remaining Euler Lagrange equations are, for the variable $\theta$, 
\begin{equation}
	I_0 \ddot{\theta} -I_0 \cos{\theta} \sin{\theta} \dot{\phi}^2+I_3\Omega \sin{\theta} \dot{\phi}+\frac{\partial V}{\partial \theta} =0\,,
\end{equation}
and for $\phi$
\begin{equation}
	I_0 \sin^2{\theta} \ddot{\phi} +\sin{\theta} \dot{\theta} (2 I_0 \cos{\theta} \dot{\phi} -I_3 \Omega)+\frac{\partial V}{\partial \phi} =0\,.
\end{equation}
In these two equations, we change variables to $x$ and $y$, make use of $\dot\Omega =0$, and make a lowest-order expansion in $x$ and $y$. As a result, the equation for $\phi$ becomes
\begin{equation}\label{eq:phiLinXy}
	\frac{I_0}{R^2} (x \ddot{y}-y \ddot{x})- \frac{I_3}{R^2} \Omega(x \dot{x} + y \dot{y}) - \frac{\partial V}{\partial x} y + \frac{\partial V}{\partial y}x =0\,,
\end{equation}
while for the one of $\theta$ we find
\begin{equation}\label{eq:thetaLinXy}
	\frac{I_0}{R^2} (x \ddot{x}+y \ddot{y})+ \frac{I_3}{R^2} \Omega(x \dot{y} - y \dot{x})+\frac{\partial V}{\partial x} x + \frac{\partial V}{\partial y}y =0\,.
\end{equation}

We can further simplify these equations by adding $x$ times equation~\ref{eq:phiLinXy} to $y$ times equation~\ref{eq:thetaLinXy} to obtain
\begin{equation}
	\frac{I_0}{R^2} \ddot{y}- \frac{I_3}{R^2} \Omega\dot{x} + \frac{\partial V}{\partial y} =0\,.
\end{equation}
Similarly, by subtracting $y$ times equation~\ref{eq:phiLinXy} from $x$ times equation~\ref{eq:thetaLinXy} we find
\begin{equation}
	\frac{I_0}{R^2} \ddot{x}+ \frac{I_3}{R^2} \Omega \dot{y}+\frac{\partial V}{\partial x}  =0\,.
\end{equation}
These last two equations can be summed up as
\begin{equation}
	\begin{pmatrix}
		\ddot{x} \\
		\ddot{y} 
	\end{pmatrix}
	=
	\frac{I_3}{I_0}\Omega 
	\begin{pmatrix}
		0 & 1 \\
		-1 & 0
	\end{pmatrix}
	\begin{pmatrix}
		\dot{x} \\
		\dot{y} 
	\end{pmatrix}
	+
	\begin{pmatrix}
		\partial_x V \\
		\partial_y V
	\end{pmatrix},
\end{equation}
which is the desired form as used in the main part of the paper.

\subsubsection*{Example for the simplification of a continuum model \ifpnas (SI)\fi}
\label{app:cont}
We motivated our approach to be relevant for more than only discrete systems. In the following we give a simple example on how it can be applied to a continuum model. The example is based on Ref.~\cite{Xiao15}, where such a reduction was performed, and we discuss how it can be cast into our classification. The system consists of a chain of dumb-bell-shaped elements, arranged in a periodic array along the axis of the dumb bells. For the details of the setup, please directly consult Ref.~\cite{Xiao15}.

The collective behaviour of the full system can be understood from a constructional point of view. Each unit cell has its eigenmodes, which get coupled to the eigenmodes of the neighboring unit cell. In general, couplings between all neighboring modes need to be considered to understand the collective behaviour of the full system such as its bandstructure. However, parts of the bandstructure can typically already be understood by taking only a few local modes into account. By an apt choice of the geometry of the unit cell \cite{Xiao15}, one can obtain a bandstructure which effectively has only two bands at around $4\,$kHz, and these two bands originate from two local modes only. Hence, if we are only interested in these two bands, we can reduce the full problem to a discrete model with only two modes per unit cell. The most general form of the $D$-matrix is then given by
\begin{equation}
	D(k)=\omega_0(k) \mathds{1}+ \vec{d}(k)\cdot \vec{\sigma} \,,\qquad \omega_0(k), d_i(k)\in\mathds{R} \,,
\end{equation}
where $\vec{\sigma} $ is the vector of Pauli matrices.

The exact coefficients depend on the details of the implementation. Nevertheless the structure of them can already be deduced from the symmetry properties of the modes and their couplings. As it turns out, one of the two modes is symmetric along the axis, while the other mode is anti-symmetric. The two modes have different eigenfrequencies and every symmetric (anti-symmetric) mode couples to its two adjacent symmetric (anti-symmetric) modes with the same coefficient due to periodicity. This implies that $\omega_0(k)=s+s'\cos(k)$ and $d_3(k)=t+t'\cos(k)$. Within a unit cell the two modes do not couple (they are eigenmodes after all), but the symmetric mode on one site couples to the anti-symmetric mode on the next site. As they have different symmetries, the coupling carries an alternating sign. From this follows that $d_1(k)=0$ and $d_2(k)=u \sin(k)$. We therefore find that
\begin{equation}
	D(k)= \omega_0(k)\sigma_0+u \sin(k)\sigma_2+[t+t'\cos(k)]\sigma_3\,.
\end{equation}

The $D(k)$ matrix has the standard $\mathcal{T} $-symmetry $U_{\mathcal{T}}=\mathds{1}\kappa $ and it has a high-frequency particle-hole and chiral symmetry
\begin{equation}
	U_{\mathcal{C} }
	=
	\begin{pmatrix}
		0 & 1 \\
		1 & 0
	\end{pmatrix}\kappa\,,
	\qquad
	U_{\mathcal{S} }
	=
	\begin{pmatrix}
		0 & 1 \\
		1 & 0
	\end{pmatrix}.
\end{equation}
Because $\Gamma=0$, the symmetries of $\sqrt{D}(k) $ get lifted to symmetries of $H(k)$ and we find that the high-frequency part of the spectrum belongs to class BDI. The $\mathds{Z} $ topological index is given by the winding number \cite{Chiu15}
\begin{equation}
	\nu=\frac{i}{2\pi } \int_{-\pi}^\pi dk \tr (q^{-1}	\partial_k q)
	=
	\begin{cases}
		\mathop{\mathrm{sign}}(t') & |t|<|t'| \\
		0 & \text{otherwise}
	\end{cases},
\end{equation}
where $q=d_3(k)-d_1(k)$. 

By connecting two chains with distinct topological coefficients, a localized mode must exist at the interface. Such a configuration has been built and the localized mode was experimentally observed in Ref.~\cite{Xiao15}.

\ifpnas 
\else

\bibliographystyle{phd-url}

\fi

\bibliography{ref,comments}

\end{document}